\documentclass[letterpaper,11pt]{article}

\usepackage{graphicx}
\usepackage[left=1.1in, right=1in, top=1in, bottom=1in]{geometry}
\usepackage{setspace}
\usepackage{booktabs}
\usepackage{xltabular}
\usepackage{multirow}
\usepackage{enumitem}
\usepackage{placeins}
\usepackage{pdflscape}
\usepackage[labelfont=bf, font=small]{caption}
\usepackage[version=4]{mhchem}
\usepackage{amsmath}
\usepackage[dvipsnames]{xcolor}
\usepackage{titlesec}
\usepackage[backend=biber,style=nature,doi=true,url=false]{biblatex}
\usepackage{hyperref}

\hypersetup{
  pdftitle={Towards Informatics-Driven Design of Nuclear Waste Forms},
  pdfauthor={Vinay I. Hegde <vhegde@citrine.io>},
  colorlinks=true,
  linkcolor=Magenta,
  citecolor=blue,
  urlcolor=blue,
}

\titleformat{\section}{\large\bfseries}{\thesection}{1em}{}

\DeclareCaptionLabelSeparator{colon}{.~}

\newcommand{\degc}{$^{\circ}$C}
\newcommand{\figref}[1]{Figure~\ref{#1}}
\newcommand{\tabref}[1]{Table~\ref{#1}}
\newcommand{\secref}[1]{Section~\ref{#1}}

\begin{document}

\begin{centering}

\textbf{\large Towards Informatics-Driven Design of Nuclear Waste Forms}\\[0.5\baselineskip]

\small{
Vinay~I.~Hegde,\textsuperscript{1}
Miroslava~Peterson,\textsuperscript{2}
Sarah~I.~Allec,\textsuperscript{1}
Xiaonan~Lu,\textsuperscript{2}
Thiruvillamalai~Mahadevan,\textsuperscript{3}
Thanh~Nguyen,\textsuperscript{3}
Jayani~Kalahe,\textsuperscript{3}
Jared~Oshiro,\textsuperscript{2}
Robert~J.~Seffens,\textsuperscript{2}
Ethan~K.~Nickerson,\textsuperscript{2}
Jincheng~Du,\textsuperscript{3}
Brian~J.~Riley,\textsuperscript{2}
John~D.~Vienna,\textsuperscript{2}
James~E.~Saal\textsuperscript{1,}\footnote{Corresponding author; email:
\href{mailto:jsaal@citrine.io}{\url{jsaal@citrine.io}}}\\[0.5\baselineskip]

$^{\rm 1}$\textit{Citrine Informatics, Redwood City, CA 94063}\\
$^{\rm 2}$\textit{Pacific Northwest National Laboratory, Richland, WA 99352}\\
$^{\rm 3}$\textit{University of North Texas, Denton, TX 76203}
}

\end{centering}

\clearpage

\begin{abstract}
\textbf{
Informatics-driven approaches, such as machine learning and sequential experimental
design, have shown the potential to drastically impact next-generation materials
discovery and design.
In this perspective, we present a few guiding principles for applying informatics-based
methods towards the design of novel nuclear waste forms.
We advocate for adopting a system design approach, and describe the effective usage of
data-driven methods in every stage of such a design process.
We demonstrate how this approach can optimally leverage physics-based simulations,
machine learning surrogates, and experimental synthesis and characterization, within a
feedback-driven closed-loop sequential learning framework.
We discuss the importance of incorporating domain knowledge into the representation of
materials, the construction and curation of datasets, the development of predictive
property models, and the design and execution of experiments.
We illustrate the application of this approach by successfully designing and validating
Na- and Nd-containing phosphate-based ceramic waste forms.
Finally, we discuss open challenges in such informatics-driven workflows and present an
outlook for their widespread application for the cleanup of nuclear wastes.
}
\end{abstract}

\section{Introduction to Nuclear Waste Forms}\label{sec:introduction}
The United States manages large volumes of nuclear wastes generated from nuclear weapons
production during World War II and the Cold War.
These wastes are primarily stored at US Department of Energy sites such as Hanford (near
Richland, Washington), Savannah River Site (near Aiken, South Carolina), Oak Ridge
National Laboratory (ORNL; in Oak Ridge, Tennessee), Los Alamos National Laboratory (in
Los Alamos, New Mexico), and the Idaho Nuclear Technical and Engineering Center (near
Idaho Falls, Idaho).
These wastes range from solids to sludges to liquids and contain, in broad terms, most
of the elements from the periodic table.
In general, the most hazardous of these wastes are high-level tank wastes at Hanford
and Savannah River which will be vitrified into borosilicate glass-based waste forms and
buried in disposal facilities.

Recent increased interest in advanced nuclear power technologies has spurred interest in
advanced nuclear fuel cycles~\cite{arostegui2019advanced}.
Some of those fuel cycles include the generation of unique waste streams such as molten
halide salts containing fission and activation products that will need to be treated for
ultimate disposal~\cite{ebert2006testing, bateman2007current, frank2017literature,
riley2019molten, riley2020electrochemical}.
Modern approaches that short circuit the traditional inefficient Edisonian
trial-and-error style of waste form development and significantly accelerate the design
of novel waste forms are in great need.

In general, options for immobilizing entire salt wastes, i.e., without any salt fraction
partitioning, are limited.
For such full-salt immobilization, the few demonstrated waste form options include
glass-bonded sodalite and tellurite glasses~\cite{ebert2006testing, riley2017glass}, all
of which exhibit low salt loading.
Alternatively, rather than immobilizing the full-salt waste, the salt can be partitioned
into different constituents to make waste form fabrication and/or partition recycling
easier.
One method for doing this is removing the halide fraction and converting the resulting
salt cations (e.g., \ce{M^{m+}}) to other chemistries (e.g., \ce{M2O}, \ce{M3PO4})
through a process called dehalogenation.

Dehalogenation can be accomplished through a variety of methods, including the
conversion of the halide salt cation in air to oxides~\cite{riley2018waste,
chong2022thermal, dong2022dechlorination}, conversion of halide salt cations
to phosphates through reactions with \ce{NH4H2PO4}~\cite{donze2000thermal,
riley2020dehalogenation, riley2021dechlorination},
\ce{(NH4)2HPO4}~\cite{riley2020dehalogenation}, or \ce{H3PO4}~\cite{siemer2012improving,
park2008stabilization}, or converting fluoride-salt wastes to a different
fluoride-containing compound that is environmentally stable (water insoluble), such as
\ce{CaF2}, for disposal~\cite{gregg2020hot}.
Any halide-containing byproducts from the dehalogenation process (e.g., \ce{H{}^{37}Cl},
\ce{NH4{}^{37}Cl}) can be captured and recycled to produce new actinide halides or
directly disposed (e.g., F).
Since the one of the key limiters of waste loading capacity of a waste form is the
halide fraction, the dehalogenated salt products (e.g., oxides, phosphates) can be
immobilized at much higher salt cation loadings than the full-salt wastes prior to
dehalogenation.


Different salt-based nuclear waste streams, such as alkali-based (e.g., Li/K/Na, Li/Be)
and/or alkaline earth-based ones, can thus undergo a dehalogenation process with the
waste cations being converted to the same anion-type compounds (e.g., oxides,
phosphates). 
These compounds then be mixed with glass-forming compounds to create a final waste form
that meets all of the necessary product criteria based on disposal requirements.
Some of the primary properties of interest for each waste composition are waste loading
(salt cation loading in wt\%), chemical durability (e.g., aqueous solubility, leach
rates), processability/manufacturability (e.g., higher melt viscosity, lower melt
temperatures), mechanical durability (e.g., compressive strength), and radiation/thermal
stability under expected radiation fields in the final product.


Historically, glass composition-property models have been developed to predict the
properties required for efficient processing and acceptable product qualities of (mainly
borosilicate) waste forms.
These models have largely been empirical fits of glass composition-property data using
single metal oxide concentrations as features.
The model development process typically involved the production and development of
painstaking, time-consuming, and empirically-created databases.
There is an opportunity to leverage the large amounts of data present in existing
chemical, physical, and thermodynamic databases to build machine learning (ML) models to
predict the waste form properties of interest for an entirely new class of waste forms,
e.g., dehalogenated phosphate ceramics or glasses.
Thus, the empirical data collection process could be side-stepped entirely or be limited
to incremental efforts in the novel chemistries and/or processes where there exist no
prior data for the ML models to train on.
Such an approach has the potential to bring about a paradigm shift in the design of
different types of novel waste forms, including relatively simple single-phase systems
(e.g., glasses, single-phase ceramics) or more complex multiphase systems (e.g.,
glass-ceramic hybrids, multiphase ceramics, and cermets).

\section{AI/ML for Nuclear Waste Immobilization: Prior Art}\label{sec:prior-art}
Approaches utilizing artificial intelligence (AI) or machine learning (ML) techniques
have enabled significant advances in many materials science problems, ranging from
the prediction of complex materials properties (e.g., superconducting critical
temperatures of complex oxides~\cite{stanev2018machine, meredig2018can}, the casting
size of metallic glass alloys~\cite{ward2018machine}) to the development of self-driving
labs~\cite{abolhasani2023role, szymanski2023autonomous}.
However, there remain certain families of materials and research questions that have not
benefited as extensively from AI/ML due to their extreme complexity, of which designing
new nuclear waste form materials is a prime example.
In the case of nuclear waste form design, difficulties in data-driven approaches
primarily arise from 
(1) a vast and complex design space (in terms of chemistry, phase, microstructure,
processing) that spans glasses, ceramics, and glass-ceramic composites, and
(2) a lack of sizeable legacy datasets for several properties a waste form must possess
to be considered viable for disposal (e.g., chemical durability, radiation stability),
especially for classes of materials that have not been previously explored.
Further, due to the cost of performing experiments with nuclear materials and the
complexity of waste form compositions, experimental data acquisition is a major
bottleneck~\cite{guerin2009materials, allen2010materials, zinkle2013materials}.


Nonetheless, there is a plethora of prior work on using AI/ML approaches to model waste
forms~\cite{morgan2022machine, hu2023data}, and we highlight a few recent reports below,
including both ceramic and glass based waste forms.
In terms of radiation effects, Pilania et al.~\cite{pilania2017using} used ML to explore
the physical factors underlying amorphization resistance in pyrochlores (\ce{A2B2O7}),
which have been extensively investigated for use in nuclear waste
forms~\cite{sickafus2000radiation, begg2001heavy, lian2003order, lian2003radiation,
lian2006effect, helean2004formation, sickafus2007radiation, lumpkin2007nature,
sattonnay2013structural, li2012role} and have been incorporated into some variants of
the SYNROC waste form~\cite{ringwood1979immobilisation}.
Here, an ML model was trained to predict the critical amorphization temperature, T$_{\rm
c}$, from simple structural features and DFT energetics.
Another critical property for nuclear waste forms is chemical durability, which is
significantly reduced by nepheline (\ce{NaAlSiO4}) precipitation during vitrification of
certain waste glasses.
To address this challenge, Sargin et al.~\cite{sargin2020data} and Lu et
al.~\cite{lu2021predicting} built ML models to predict nepheline crystallization
behavior from glass composition, comparing several different algorithms and achieving a
reasonable classification accuracy.
However, due to the data acquisition challenge for nuclear waste forms, we note that the
dataset sizes for the aforementioned models is relatively low for ML (on the order of
100--1000 data points), which may limit generalizability and extrapolability.

Similarly, several previous efforts focused on developing ML models for predicting
properties relevant for glass waste forms, including glass
density~\cite{hu2020predicting}, viscosity~\cite{cassar2021viscnet},
durability~\cite{krishnan2018predicting, han2020machine, brauer2007solubility}, glass
transition temperature~\cite{cassar2018predicting}, thermal
expansion~\cite{mastelini2022machine},  mechanochemical
wear~\cite{qiao2021applicability}, and Young's modulus~\cite{yang2019predicting}.
While many of these models were trained to predict a single property on a specific
family of glasses (e.g., silicates), a recently developed multi-task deep neural network
model, GlassNet, has been trained predict 85 various glass properties on the entire
SciGlass database~\cite{sciglass} with reasonable accuracy on most of the modeled
properties~\cite{cassar2023glassnet}.

In addition to the direct modeling of waste form properties, the development of neural
network interatomic potentials (NNIPs) trained on first-principles density functional
theory (DFT) energetics of waste form-relevant systems has made significant strides.
For example, NNIPs have been applied to the modeling of molten salts, enabling the
accurate prediction of structural and dynamical properties at normal operating
conditions, high-temperature–pressure conditions, and in the crystalline solid
phase~\cite{lam2021modeling, li2021development}.
Additionally, Byggm\"{a}star et al. developed a Gaussian Approximation Potential (GAP)
to study radiation damage in W~\cite{byggmastar2019machine}, and Ghosh et al. developed
an NNIP to study Cs incorporation into hollandite
(\ce{A2B8O16})~\cite{ghosh2020utilization}.

While significant advances have been made in applying AI/ML methods toward modeling
nuclear waste forms, approaches that utilize such methods for the design of entirely novel
waste forms are not well established.
Some recent works have explored the use of AI/ML to design optimal waste glass
compositions~\cite{gunnell2022machine, lu2023glass}, but a unified framework for an
informatics-driven design, especially in previously-unexplored chemistries encompassing
ceramics, glasses, and other types of waste forms is lacking.

In the following sections, we present a generally-applicable, closed-loop, iterative
design framework that effectively combines AI/ML approaches with physics-based
simulations and experiments to design novel waste forms.
As a topical use-case, we present the design of new phosphate-based \textit{ceramic}
waste forms as part of an ongoing project under the Advanced Research Projects
Agency-Energy (ARPA-E) ONWARDS program, with the following initial target properties:
(1) {\textgreater}20\% waste cation mass loading,
(2) {\textless}400~cm$^3$ waste form volume/100~g salt, and
(3) {\textgreater}30~MPa compressive strength.
For each task in the presented iterative design framework, we discuss available methods
and tools, and best practices to use them, with illustrative examples from the
phosphate-based waste form design problem.
We also present a single end-to-end pass (i.e., the first iteration) through the design
process for the same problem, resulting in successful synthesis of phosphate waste forms
that satisfy all the target criteria listed above.

\section{A System Design View of Nuclear Waste Forms}\label{sec:system-design}
A system design approach primarily uses the well-established
processing-structure-property-performance (PSPP) paradigm to understand the underlying
materials problem~\cite{olson1997computational, kuehmann2009computational}.
Namely, the ways in which a material is synthesized and processed (including the various
steps and their respective parameters and conditions) determines its structure, both on
a microscopic level (i.e., the phases that are formed, their atomistic structure, etc.)
as well as on a macroscopic level (i.e., the grain structure, level of porosity and
other defects, etc.). 
The overall structure of a material, in turn, determines its properties, which, in turn,
determine its performance in the target application.
This system design approach can be encapsulated in a so-called ``system design chart''
where individual components of the PSPP systems are connected by causal links or edges
in the forward direction (i.e., from processing to structure to properties to
performance), which often capture domain knowledge for the problem.
Note that several processing steps can affect multiple structural features of a
material, and several structural features can contribute to the properties of the
overall material, and so on.
Drawing up a system design chart for the materials problem at hand should therefore be
the first step in any non-trivial materials design endeavor.

\begin{figure}[!htb]
  \centering
  \includegraphics[width=0.9\textwidth]{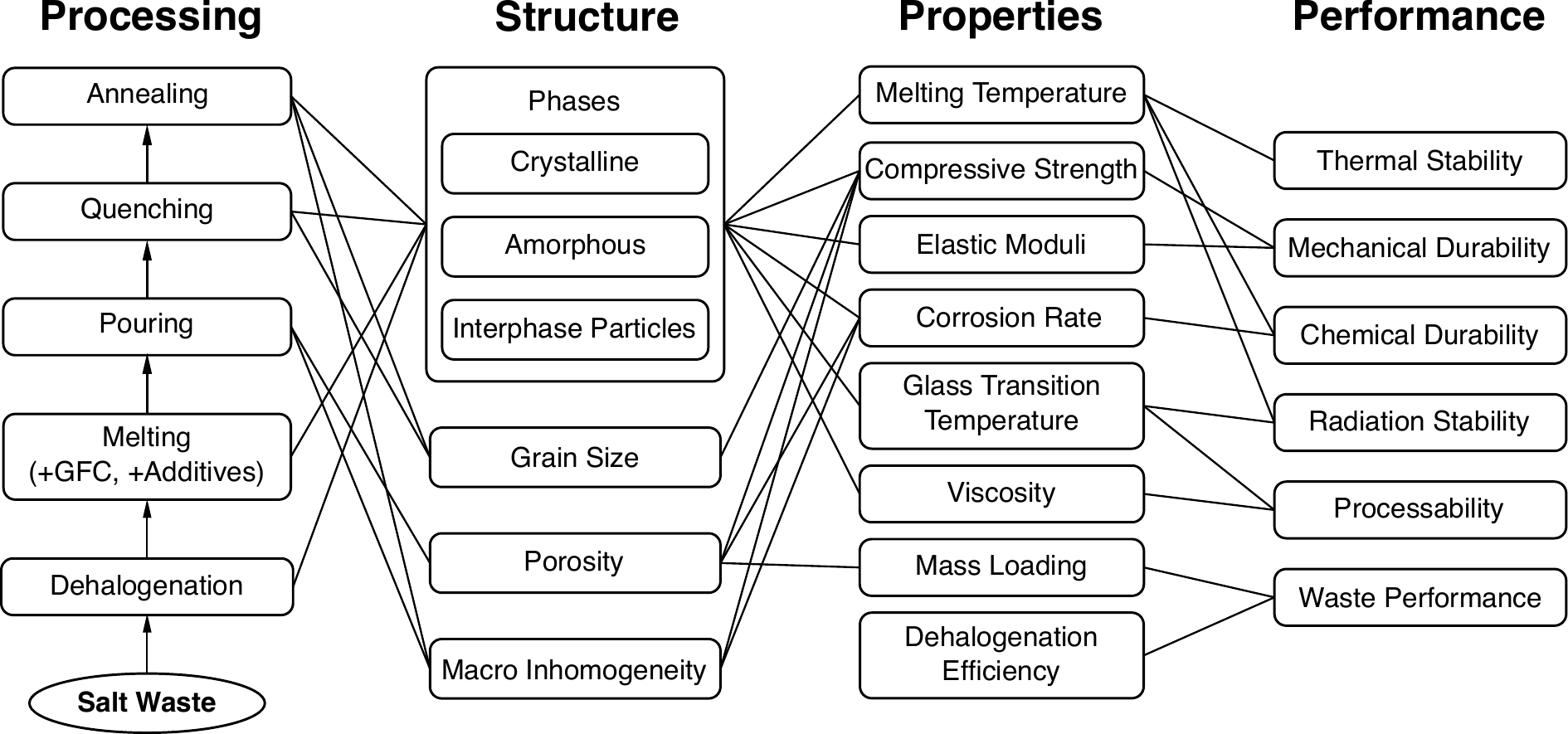}
  \caption{
    A sample system design chart for nuclear waste form design.
  }\label{fig:system-design}
\end{figure}

An example system design chart for the recently-proposed phosphate glass waste
forms~\cite{riley2020dehalogenation} is shown in \figref{fig:system-design}.
Here, a typical processing step can involve dehalogenation of the salt waste via
reagents such as \ce{NH4H2PO4}, followed by melting with glass-forming chemicals (GFCs)
such as \ce{Fe2O3}, and eventual quenching and/or slow cooling of the melt.
The waste form resulting from such a process can be crystalline, glassy, or a mixture of
both, i.e., various crystalline phases embedded in a glassy matrix.
In addition, depending on other processing steps such as sintering or hot isostatic
pressing (HIP), the waste form may have porosities ranging from 2\% to 30\%.
The properties of interest of the resulting waste forms can then be obtained by direct
experimental measurement during one of the processing steps (e.g., melting temperature
and viscosity during the melting step) or once the processing is complete (e.g., waste
salt cation loading, volume, and compressive strength of the final solid waste form).
Finally, within limitations of available experimental facilities, the waste form can be
tested to determine its actual performance in near-operation conditions (e.g., chemical
durability in aqueous conditions).

\section{A System Design-Informed Sequential Learning Framework}\label{sec:sl}
Sequential learning (or active learning) (SL) is an iterative approach for the design of
novel materials via data-driven models.
A typical SL workflow consists of the following iterative loop: 
(1) machine learning (ML) models are trained on existing materials property data,
(2) these ML models are then used to make property predictions in an as-yet unexplored
design space,
(3) the ML predictions, and uncertainties in such predictions, are used to identify the
most-promising candidate material(s) to evaluate next,
(4) the objective evaluation (via simulations and/or experiments) of the selected
candidate(s) is performed, and
(5) the loop iterates to augment the training data with results from the objective
evaluations, retrain and/or refine the ML models, make new candidate predictions, and so
on.
SL workflows have been previously shown to significantly accelerate the discovery of new
materials, ranging from small molecules to catalysts to semiconductors and
others~\cite{ling2017high, graff2021accelerating, annevelink2022automat,
antono2020machine, kavalsky2023much}, as well as to rapidly optimize manufacturing and
processing parameters~\cite{fong2021utilization, ling2018machine}.
 
\begin{figure}[!htb]
\includegraphics[width=\textwidth]{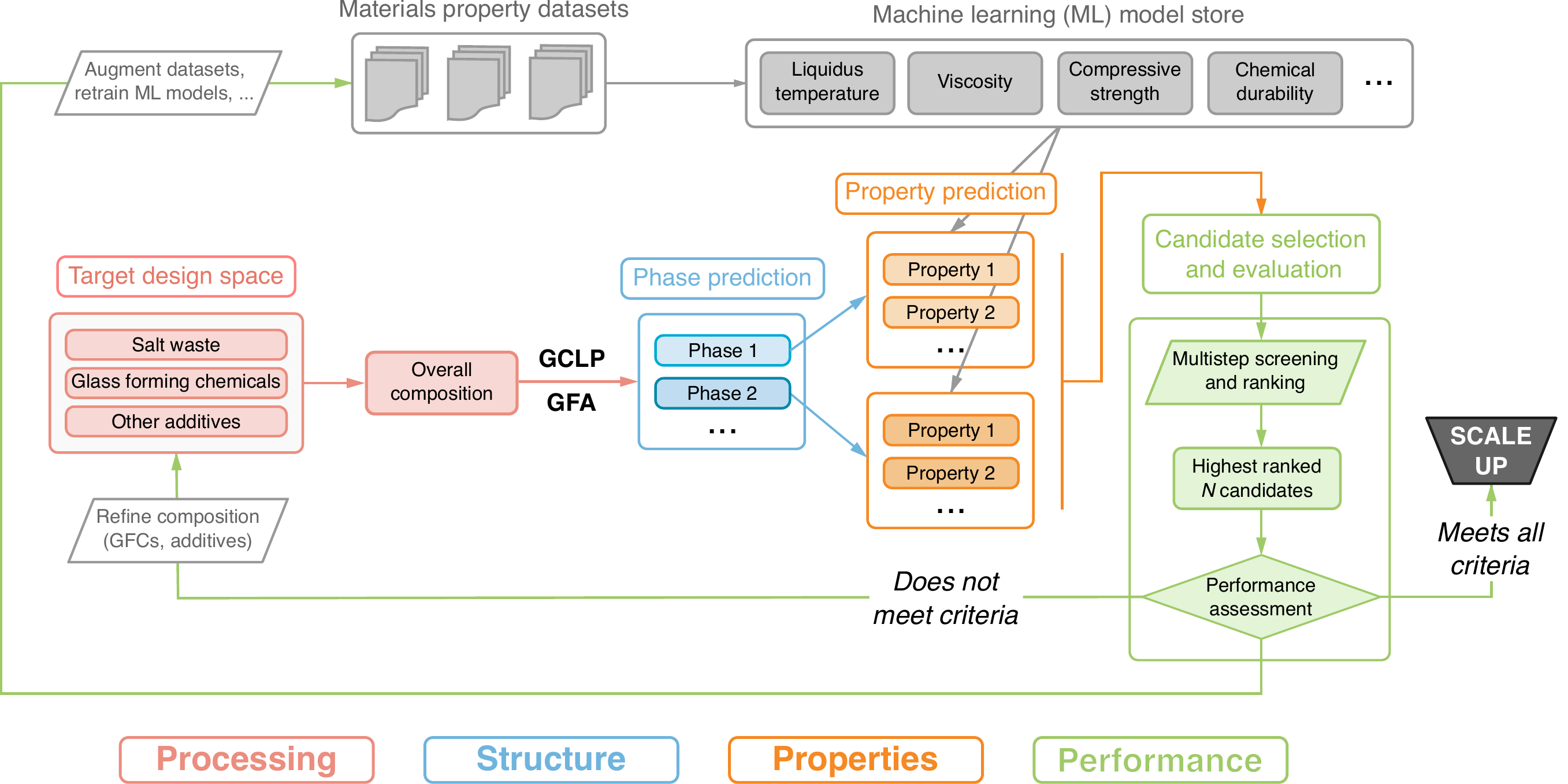}
\caption{
  A sample sequential learning driven workflow for the design of novel waste forms,
  informed by the underlying system design chart.}\label{fig:sl}
\end{figure}

The construction of such an SL workflow for the materials design problem at hand can
greatly benefit from being informed by the underlying system design chart.
An example of a system design-informed SL workflow for phosphate waste forms is shown in
\figref{fig:sl}.
An effective translation from the system design chart to a closed-loop SL workflow
requires:
(1) an initial data generation or collection effort upfront, ideally
pertaining to the design spaces of interest,
(2) various aspects of processing, structure, and properties to be adequately captured
in the representation or schema used to store materials data,
(3) physics-based simulations or ML models that can predict both (a) the mixture of
phases and their structure in the final resulting waste forms based on their initial
compositions and relevant processing conditions, and (b) the target waste form
properties of interest, with robust uncertainty estimates, and
(5) a scheme to filter and rank candidates in the target design space that takes into
account model predictions and uncertainties, as well as any other application-based
constraints not previously accounted for.
In each of the above steps, incorporating domain knowledge and expertise can have a
significant impact on the effectiveness of the SL workflow; this is further discussed in
the context of each individual SL task below.

\subsection{Dataset construction}\label{ssec:dataset_construction}

The first step in an SL workflow is dataset construction.
This can involve
(1) \textit{data collection} from various sources such as existing materials databases,
published materials property datasets, and individual or small sets of data scattered
across literature,
(2) \textit{data generation} using relatively inexpensive physics based simulations (see
\secref{ssec:physics_sim}) and/or a small set of baseline experimental measurements, and
(3) \textit{data engineering} of the collected and/or generated data.
The latter step encompasses data curation (e.g., outlier removal, imputation of missing
values), data fusion (e.g., combining data for a given set of materials from different
experimental measurements or simulations), and other related post-processing.

What datasets should one construct for their effective use in waste form design?
The target datasets should be informed by the underlying system design chart.
That is, datasets should be constructed for the target materials properties that
correspond to the waste form performance criteria.
This identification of materials properties data that correlates with waste form
performance is non-trivial and requires significant domain knowledge integration
for extracting maximum utility.
We provide a sample list of openly-available datasets of materials properties and
related performance criteria in \tabref{tab:datasets}, some of which were used for
phosphate waste form design in this work.
Note that while a 1:1 correspondence between materials property and performance metric
is ideal (e.g., leach rate [property] $\Rightarrow$ chemical durability [performance]),
ML models can leverage indirectly-related data (e.g., cohesive energy [property]
$\Leftrightarrow$ melt temperature [property] $\Rightarrow$ processability
[performance]) via approaches such as transfer learning~\cite{hutchinson2017overcoming,
yamada2019predicting, jha2019enhancing, cubuk2019screening, gupta2021cross,
chen2021atomsets} (see \secref{ssec:ml}).

{\small
  \begin{xltabular}[!htb]{\linewidth}{llll}
  \caption{
    A few sample open materials property datasets used to optimize the corresponding
    waste form performance; here, used to design phosphate-based waste forms.
  }\label{tab:datasets} \\
  \toprule
  \textbf{Performance metric} & \textbf{Material property} & \textbf{Dataset source} & \textbf{Dataset size} \\

  \midrule
  \multirow{2}{*}{Mass loading} & Formation energy & Materials Project & $\sim$140,000 \\
                                & Thermodynamic stability & Materials Project & $\sim$140,000 \\
  \midrule
  Mechanical durability & Bulk modulus & Materials Project & $\sim$5,640 \\
  \midrule
  \multirow{2}{*}{Thermal stability} & Melting temperature & Literature & $\sim$250 \\
                                     & Liquidus temperature & SciGlass & $\sim$45,280 \\
  \midrule
  \multirow{2}{*}{Processability} & Glass transition temperature & SciGlass & $\sim$91,650 \\
                                  & Viscosity & SciGlass & $\sim$6,290 \\
  \midrule
  Chemical durability & Cohesive energy & Materials Project & $\sim$140,000 \\
                      & Solubility & IUPAC-NIST & $\sim$670 \\
                      & Leach rate & ALTGLASS & $\sim$2,400 \\
  \bottomrule
\end{xltabular}

}

Lastly, while the dataset construction task is often overlooked in favor of exploring
more sophisticated ML algorithms, improving data quality and quantity is perhaps the
most critical enabler of the eventual success of an informatics-driven approach to
materials design. A recent ``renaissance'' towards data-centric AI (in contrast to
model-centric AI) across several other fields is ongoing~\cite{zha2023data,
hamid2022model, bartel2021data}.

\subsection{Physics-based simulations}\label{ssec:physics_sim}

Physics-based simulations, such as those based on first-principles density functional
theory (DFT), molecular dynamics (MD), calculation of phase diagrams (CALPHAD), and
other approaches can be used to perform ``computational experiments'', at various time
and length scales, that can both augment datasets of materials properties as well as
provide faster feedback within an SL design loop.
We discuss some of the more widely-used physics-based modeling approaches below, with a
focus on how they can be integrated within an SL workflow.


\subsubsection{Ab initio atomistic simulations}\label{sssec:dft}

\textit{Ab initio} density functional theory (DFT) and related methods can be used to
calculate a variety of properties of a material based only on its crystal structure,
including but not limited to formation energy, bulk density, elastic moduli, electronic
band structure and gap, and others.
Due to some underlying approximations in the approach (e.g., the exchange-correlation
functional in DFT), the accuracy of a DFT calculation can vary significantly based on
the target material/property and its applicability is often limited to materials with
unit cells consisting of not more than hundreds of atoms.
Nonetheless, these calculations are an invaluable tool to generate initial training data
for ML, validate ML-predicted properties of a waste form candidate, improve the
description of a waste form composition, etc.\ within an SL workflow.
As an example, below we highlight how a DFT-calculated property (compound formation
energy) can be used to estimate the ground state mixture of phases from the overall
composition of a waste melt, enabling ML property predictions for those phases.

\begin{figure}[!htb]
\centering
\includegraphics[width=\textwidth]{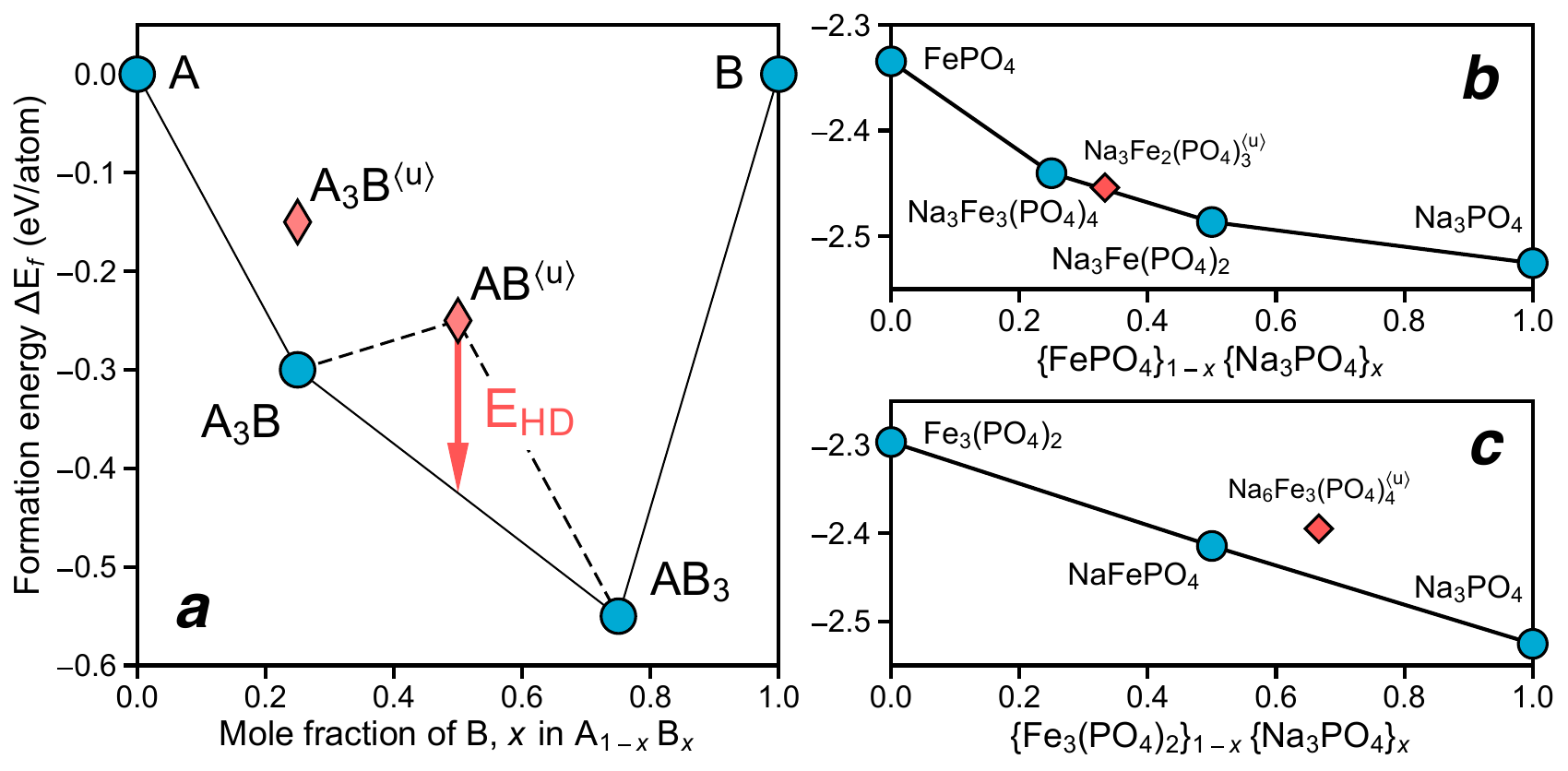}
\caption{
  (a) The convex hull of formation energy in a model A-B system. Blue circles indicate
  stable phases (``on the hull''), and red diamonds ($\langle u \rangle$) indicate
  phases that are metastable or unstable (``above the hull''). Under perfect
  thermodynamic equilibrium conditions, the unstable/metastable phases will decompose
  into a mixture of stable phases. For example, AB is predicted to decompose into a
  mixture of A$_3$B + AB$_3$, and its distance to the convex hull (``E$_{\rm HD}$'') is
  an indicator of the thermodynamic drive for the decomposition (or relatedly,
  likelihood of its experimental synthesis~\cite{kim2017experimental,
  ma2017computational}).
  (b, c) Slices of the convex hull computed using DFT-calculated formation energies in
  the Na-Fe-P-O chemical system, with sodium and iron phosphate (3+ and 2+,
  respectively) end members.
}\label{fig:hull-and-gclp}
\end{figure}

An analysis of the convex hull of compound formation energies as a function of phase
composition can be used to identify the thermodynamically stable phases in a given
chemical space~\cite{barber1996quickhull, kirklin2016high, wang2018crystal,
hegde2020phase, aykol2019network}.
Combined with linear programming approaches, a convex hull of a chemical space can
rapidly predict compounds that are stable/synthesizable as well as the ground state
mixture of phases that can be expected to form under thermodynamic equilibrium
conditions at any given composition in that chemical space (see
\figref{fig:hull-and-gclp}a)~\cite{akbarzadeh2007first}.

The grand potential $\phi$ of a collection of phases in a chemical space is given by
\begin{equation}\label{eqn:gclp}
  \phi(\boldsymbol{x}, \boldsymbol{\mu}) = \sum_i x_i G_i - \sum_j \left( \mu_j \sum_i x_i C_{i,j} \right)
\end{equation}

where $x_i$ is the relative amount of phase $i$, $\mu_j$ is the chemical potential of
element $j$, and $C_{i,j}$ is the fraction of element $j$ in phase $i$.
The ground state mixture of phases (i.e., the $x_i$) can be determined by minimizing
$\phi$ with respect to $\vec{x}$ and $\vec{\mu}$, provided the free energies $G_i$ and
the list of stable phases in the chemical system (the $C_{i,j}$ terms) are defined.
Typically, tools such as high-throughput DFT are used to calculate the energies of the
various phases in the chemical space of interest in a self-consistent fashion, and the
phase free energies are approximated by their respective 0~K formation energies.
Although using 0~K energies from a computational tool such as DFT ignores
finite-temperature effects, kinetic effects, and other factors, and can result in
differences between predictions and experimental observations (e.g., see two slices of a
DFT-computed Na-Fe-P-O convex hull in \figref{fig:hull-and-gclp}b--c, where two
experimentally-reported phases, \ce{Na3Fe2(PO4)3} and \ce{Na6Fe3(PO4)4}, are predicted
to be unstable), this approach is still invaluable for estimating ground state mixture
of phases with reasonable accuracy~\cite{kim2017experimental, ma2017computational},
particularly for systems that are not experimentally explored or fully described within
the CALPHAD approach (see \secref{sssec:calphad}).

\subsubsection{Semi-empirical atomistic simulations}\label{sssec:md}

Interatomic potential based atomistic simulations, e.g., classical molecular dynamics
(MD) and Monte Carlo (MC) methods, are computationally much more efficient and can deal
with much larger systems (up to millions or billions of atoms) and longer time scales
(micro to milliseconds) than \textit{ab initio} methods. 
Such semi-empirical atomistic simulations have been successfully used to study
multicomponent glass structures, radiation effects in metals, ceramics and glasses, and
to study the effect of microstructure on mechanical
properties~\cite{rimsza2014structural, ren2016structural, du2007erbium}.

The fidelity of an MD simulation strongly depends on the type and quality of the
underlying interatomic potential (IAP), making the development of an IAP and its
parameterization a significant portion of the overall simulation effort.
For example, the development of IAPs for the modeling of borosilicate glasses, a
widely-used nuclear waste glass, has been a challenge due to the strong dependence of
boron coordination on thermal history and composition, which in turn affects properties
such as mechanical strength and corrosion rate.
However, recent efforts in IAP development have led to successful simulations of
borosilicate glasses across a wide range of compositions where boron coordination is in
good agreement with experimental observations~\cite{deng2019development,
deng2016development, deng2022borosilicate}.
Also, the development of efficient ``reactive'' potentials allow for simulations of
reactions of aqueous solution with glasses~\cite{rimsza2022simulations,
rimsza2017interfacial, mahadevan2020hydration, mahadevan2022development,
kalahe2023composition}.

\begin{figure}[!htb]
\centering
\includegraphics[width=\textwidth]{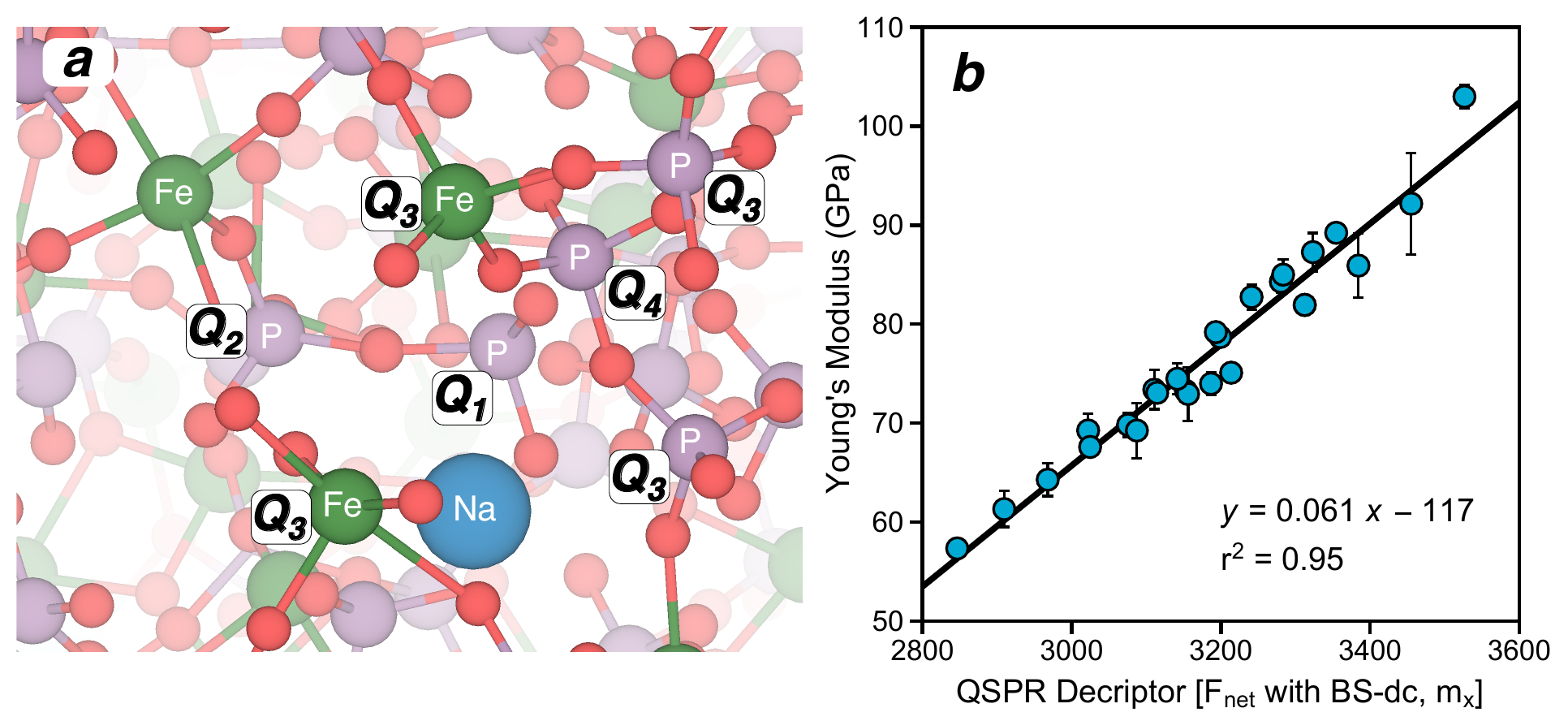}
\caption{
  (a) Visualization of part of the atomic structure of a sodium iron phosphate glass
  with composition \ce{5\%Na2O-35\%Fe2O3-60\%P2O5}, showing the various $Q_n$ units in
  the glass network (where $n$ is the number of bridging oxygen atoms (red spheres)
  connected to network formers, \ce{Fe2O3} and \ce{P2O5}), and (b) linear correlation
  between Young's modulus and QSPR-based network strength descriptor, F$_{\rm net}$,
  calculated using bond strength of diatomic cations (``BS-dc'') and a multiplicative
  factor (``m$_{\rm x}$'')~\cite{lusvardi2009quantitative, du2021predicting}. The blue
  circles are MD-calculated Young's moduli values for various Na-Fe-P-O glass
  compositions (the error bars indicate variance in the calculated moduli across
  different structures at the same composition), and the solid black line shows a linear
  fit with an $r^2$ score of 0.95.
}\label{fig:md_structure_qspr}
\end{figure}

Further, these large-scale atomistic simulations enable so-called quantitative
structure-property relationship (QSPR) analysis, linking properties of a material to its
atomistic structural features.
QSPR analysis has been applied to a wide range of properties such as glass transition
temperature, dissolution rate, Young's modulus, and hardness of glass materials,
correlating trends in these properties to descriptors derived from structural features
such as bond angles, bond energies, coordination numbers, network connectivity (e.g.,
see \figref{fig:md_structure_qspr}a for a visualization of a iron phosphate glass,
showing various $Q_n$ units, where $n$ is the number of bridging oxygen atoms connected
to network formers $Q$, \ce{Fe2O3} and \ce{P2O5})~\cite{kalahe2024insights}.
A simple statistical model (e.g., linear or multilinear regression) can then be used to
identify relationships between such descriptors and the target property of interest
(e.g., see \figref{fig:md_structure_qspr}b for an example of modeling Young's modulus as
a linear function of a glass network strength descriptor, F$_{\rm
net}$~\cite{lusvardi2009quantitative, du2021predicting}).

\subsubsection{Thermodynamic modeling}\label{sssec:calphad}

The Calculation of Phase Diagrams (CALPHAD) approach is a commonly used thermodynamic
formalism, along with related numerical methods, to assess phase equilibria in materials
under various conditions.
Within the CALPHAD approach, thermodynamic properties (e.g., Gibbs energy, specific
heat) of phases in a system are described using mathematical models with adjustable
parameters, which are then optimized by fitting to all available thermochemical
information about phases/sub-systems to arrive at a consistent description of the
multicomponent system of interest.

\begin{figure}[!htb]
\includegraphics[width=\textwidth]{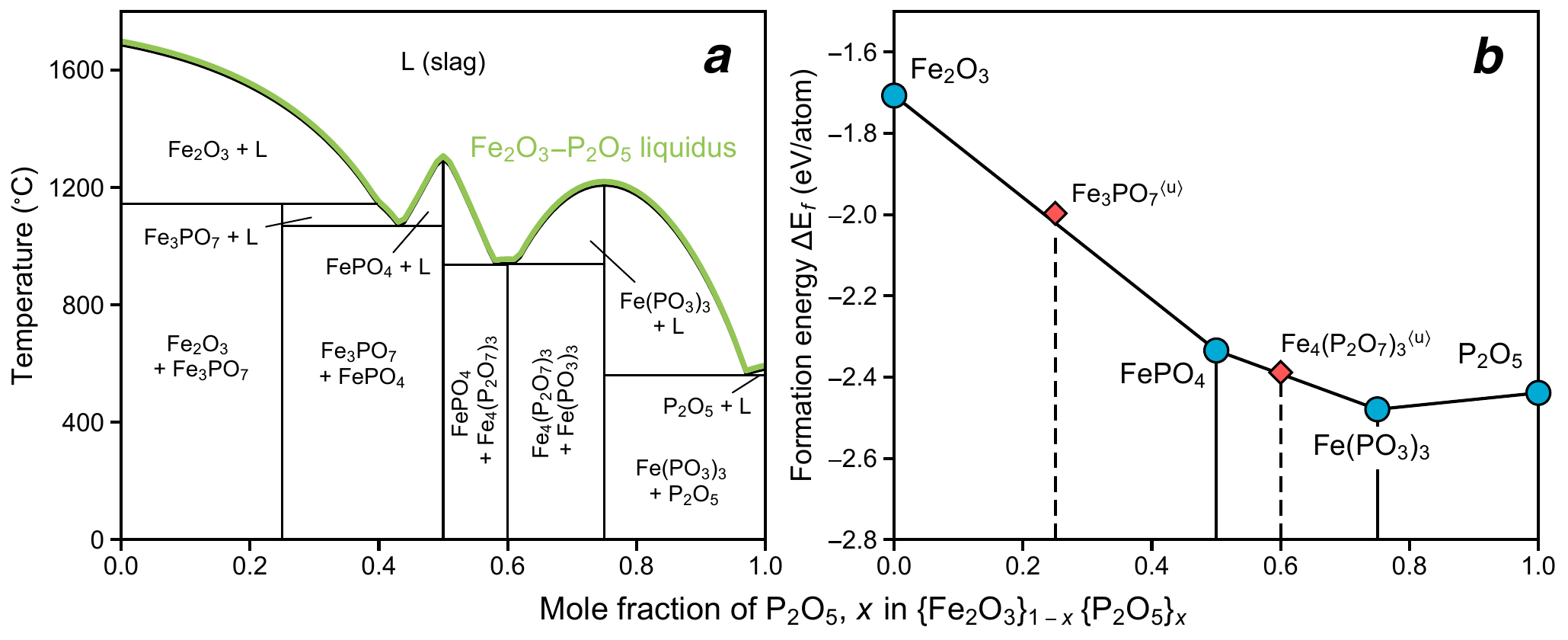}
\caption{
  The \ce{Fe2O3-P2O5} phase diagram (a) computed using the CALPHAD approach (the
  liquidus curve is highlighted in solid green), and (b) calculated using \textit{ab
  initio} DFT (blue circles indicate ``on the convex hull'' stable phases and red
  diamonds indicate ``above the convex hull'' 0~K metastable phases).
}\label{fig:phase_diagrams}
\end{figure}

The CALPHAD method can be used to study the typically multicomponent, multiphase nuclear
waste materials, e.g., calculate phase diagrams of the waste form systems (e.g., the
\ce{Fe2O3-P2O5} iron phosphate system, shown in \figref{fig:phase_diagrams}),
equilibrium and non-equilibrium phase evolution during solidification of a molten salt
mixture, estimating transformation temperatures between waste-relevant phases, waste
salt or waste element solubility limits, and so on.

While CALPHAD-based approaches can be used to study the behavior of waste-relevant
systems under non-ideal conditions (e.g., as a function of temperature, pressure,
chemical potential, pH), they are often limited by the availability of fully-assessed
thermodynamic databases for the systems of interest.
While \textit{ab initio} methods can be used to calculate the properties (e.g.,
formation energy) of novel phases from scratch, such data may suffer from limitations
related to ``ideal conditions'', as discussed in \secref{sssec:dft}.
For example, we compare the 0~K phase diagram calculated using DFT with that from
CALPHAD for the \ce{Fe2O3-P2O5} system in \figref{fig:phase_diagrams}. 
Note that two of the experimentally observed phases, \ce{Fe3PO7} and \ce{Fe4(P2O7)3},
described correctly by CALPHAD are predicted to be metastable at 0~K by DFT.
Further, for systems with missing or partial CALPHAD assessments, there are no currently
available tools that enable a seamless integration of such \textit{ab initio} data with
existing thermodynamic databases to extend their capabilities to the partially
assessed/unassessed system (e.g., the \ce{Na2O-Fe2O3-P2O5} system, discussed in
\secref{ssec:expt}).

\subsection{Machine learning models}\label{ssec:ml}

The various waste form-relevant property datasets described in
\secref{ssec:dataset_construction} can be used as input to train ML models that can
predict those properties of novel waste forms.
While sizeable research efforts have gone into ML algorithms and models for predicting
materials properties, there are a few considerations to keep in mind specifically
towards their effective applicability in an SL workflow for waste forms.
First, the crystal structure of a novel waste form is often unknown or yet to be
characterized.
Even though ML models that are built using crystal structure-based descriptors as input
show higher accuracy in predicting materials properties~\cite{dunn2020benchmarking},
this information is often \textit{a priori} unavailable.
Thus, ML models that can predict waste form properties purely based on the melt
composition alone tend to be of higher value than those that require crystal structure
as input.
Several approaches have been developed to generate physical descriptors based on
chemical composition of a material (``featurization''), e.g.,
Magpie~\cite{ward2016general}, Deml~\cite{deml2016predicting},
Matscholar~\cite{weston2019named}, MEGNet~\cite{chen2019graph}, including off-the-shelf
software tools with implementations of such approaches, e.g.,
matminer~\cite{ward2018matminer}, that can be used to featurize waste form compositions.

Second, the type of AI/ML model to build, e.g., based on ensemble approaches such as random
forests (RF) and gradient-boosted trees, or deep neural net (NN) approaches such as
ElemNet~\cite{jha2018elemnet}, Roost~\cite{goodall2020predicting},
CrabNet~\cite{wang2021compositionally}, depends on the target property and the size of the
available data.
Tree-based models have been shown to generally outperform deep learning models on
tabular data, especially for small-to-medium sized datasets typical of materials
properties (10$^2$--10$^3$ examples)~\cite{grinsztajn2022tree}.
While NN-based models are useful for certain larger property datasets, e.g., CrabNet for
predicting the formation energy, or GlassNet to predict the glass transition temperature
from an input composition), we urge researchers to test ``simpler'' ML models using
off-the-shelf tools such as scikit-learn~\cite{scikit-learn} and lolo~\cite{lolo}, as
these can often be more interpretable and provide superior extrapolation
performance~\cite{Muckley_2023}.

\begin{figure}[!htb]
\centering
\includegraphics[width=\textwidth]{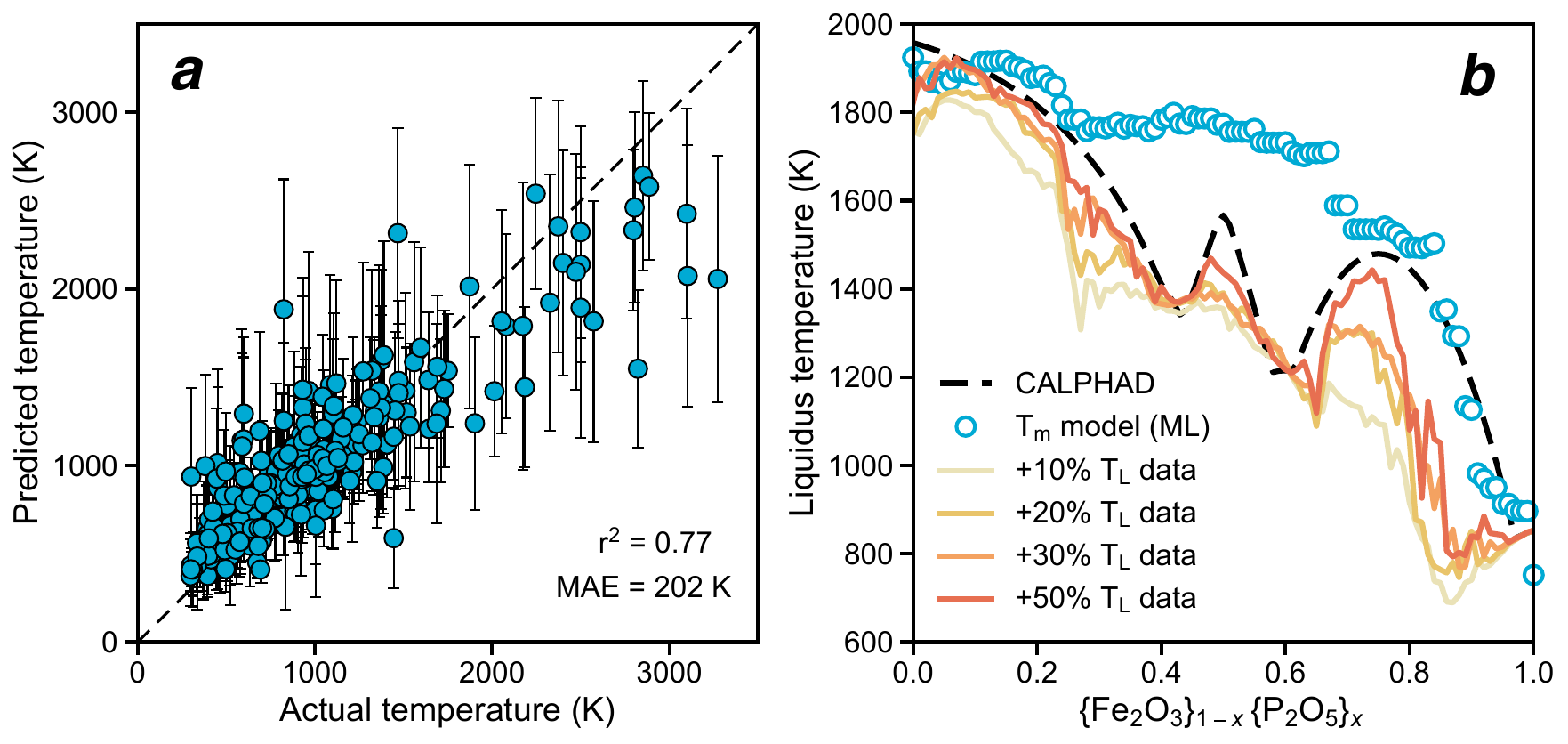}
\caption{
  (a) An actual-vs-predicted parity plot for an ML model trained in predict congruent
  melting temperatures of stoichiometric compounds, evaluated using 5-fold random
  cross-validation. 
  (b) Performance of the melting temperature model in predicting liquidus temperatures
  of a waste form-relevant target system (here, \ce{Fe2O3-P2O5}). The dashed black lines
  represent ``ground-truth'' liquidus temperatures from CALPHAD, and the blue circles
  are liquidus values predicted by the melting temperature (T$_{\rm m}$) model.
  The yellow-to-red solid lines show predictions from an ML model that is trained on a
  dataset with increasingly higher amounts of the target liquidus data (T$_{\rm L}$) in
  the training set, tested on the held-out liquidus data.
  These results show that the performance of the model, especially in the two deep
  eutectic-related compositions of 40\% and 60\% \ce{P2O5} can be improved significantly
  augmenting the training dataset with even a small sampling (10--20\%) of the target
  design space.
}\label{fig:ml}
\end{figure}

Third, we underline the importance of assessing the performance of ML models using
evaluation methods and metrics that are directly relevant to the task at hand.
Typically, ML model accuracies are reported using metrics such as r$^2$ values or mean
absolute errors (MAEs) evaluated in random cross-validation or a random
train/validation/test split of the available data, which can be informative of
within-distribution generalization but not of out-of-distribution performance. 
For example, an ML model for predicting melt temperatures trained on chemical
composition based Magpie features~\cite{ward2016general} and a dataset of congruent
melting temperatures of stoichiometric compounds has a reasonable accuracy (MAE of
$\sim$200~K, consistent with prior work~\cite{seko2014machine}) in random
cross-validation (\figref{fig:ml}a).
However, the model fails to capture the liquidus temperature trends in a target design
space of the \ce{Fe2O3-P2O5} system, especially the two deep eutectic regions around
40\% and 60\% \ce{P2O5} content, where the prediction errors are higher than 400--450~K
(see \figref{fig:ml}b).
In short, using a test strategy (e.g., leave-one-cluster-out [LOCO] cross-validation for
estimating extrapolative performance~\cite{meredig2018can}) that is representative of
the actual environment that the ML model will be eventually used in, is critical for a
true assessment of its performance.
Further, an accurate evaluation of a model is crucial to improving it, e.g., overcoming
data scarcity with strategies such as transfer learning~\cite{hutchinson2017overcoming,
yamada2019predicting, jha2019enhancing, cubuk2019screening, chen2021atomsets,
gupta2021cross} or prioritizing targeted data collection (see \figref{fig:ml}b, which
shows how adding even a small fraction [10--20\%] of data from the target design space
of \ce{Fe2O3-P2O5} enables the ML model to learn the liquidus curve more effectively
than from congruent melting temperatures alone).

Lastly, while most of the focus of ML-based modeling of materials properties has been
towards developing more accurate models, the efficacy of an ML model within an SL
workflow goes beyond simply the model accuracy~\cite{Borg_2023}.
In other words, it is indeed possible to have performant SL workflows with
lower-accuracy ML models.
Further, robust uncertainty estimates along with ML model predictions are crucial not
only to calibrate user confidence in the model predictions but also for ranking and
selecting candidates within an SL workflow (see \secref{ssec:candidates}).

\subsection{Candidate ranking and selection}\label{ssec:candidates}

Once ML models for target properties have been trained, the next task in the SL workflow
is to rank and select candidates in a target design space for objective evaluation.
The target design space is defined by the target application, and can be identified by
the underlying system design chart.
As discussed earlier (see \secref{sec:system-design}), for phosphate-based nuclear waste
forms, the design space is often bound by the composition of the liquid melt, which in
turn is defined by the waste stream, the reagents used for dehalogenation, glass-forming
chemicals (GFCs), and other additives added during the processing step.
From a well-defined design space, candidates can be generated via simple enumeration
(e.g., a uniform grid of compositions within a Gibbs triangle for a ternary design
space).
This enumerated list of candidates can be further filtered down using domain knowledge
and/or outputs of simulations as an intermediate step.

\begin{figure}[!htb]
\centering
\includegraphics[width=\textwidth]{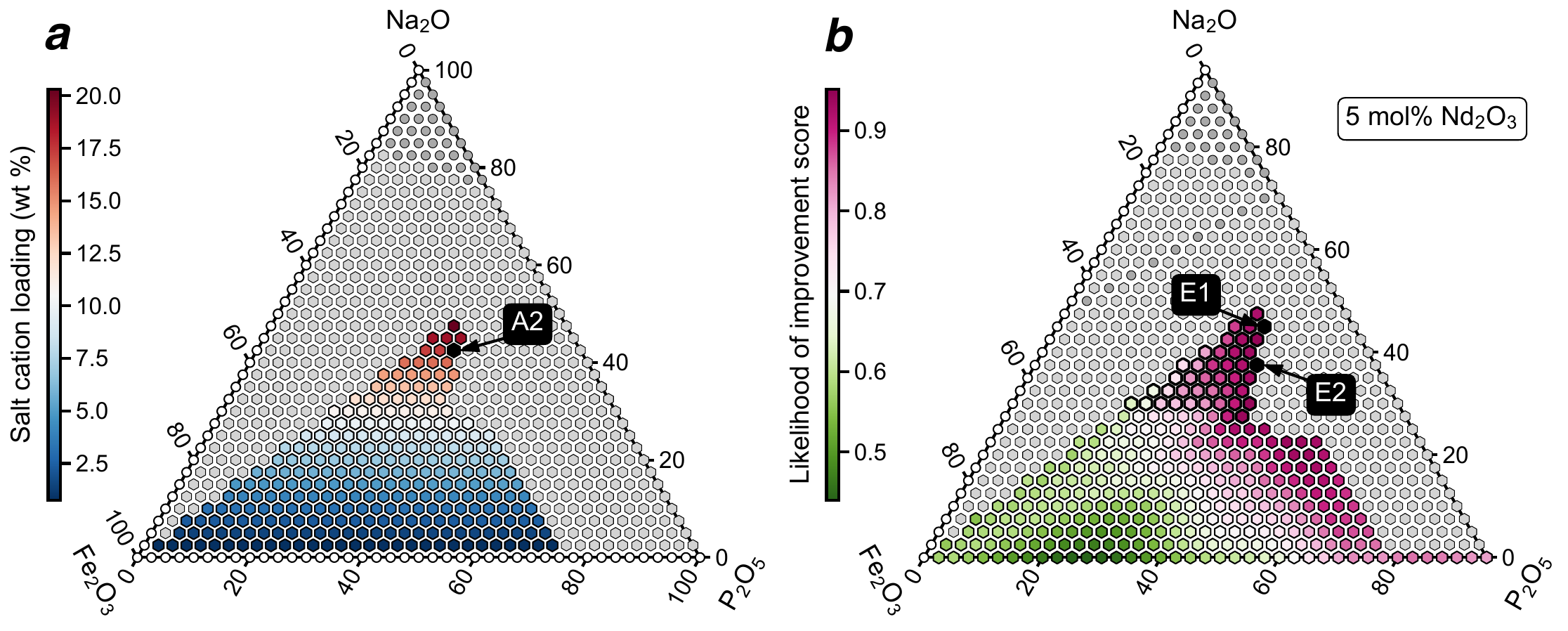}
\caption{
  Candidate generation, ranking, and selection in two sample waste form design spaces:
  (a) \ce{Na2O-Fe2O3-P2O5} and (b) \ce{Na2O-Nd2O3-Fe2O3-P2O5} (with a constant 5 mol\%
  \ce{Nd2O3}).
  In both design spaces, a significant fraction of the candidate compositions are
  filtered out: with no salt waste or phosphate phases (open circles), with unreacted
  waste (dark grey filled circles), with binary alkali phases (light grey filled
  hexagons).
  Viable candidate compositions are shown as hexagons colored according to (a) salt
  cation loading in wt\%, and (b) a scaled score that quantifies the likelihood of
  improvement simultaneously over \textit{all} target metrics over reasonable baseline
  values; in the latter panel, candidate compositions with salt cation loading $>$20
  wt\% are shown as hexagons outlined in bold.
  The candidates selected for experimental synthesis and validation from the two design
  spaces, (a) A2 and (b) E1, E2, are annotated.
}\label{fig:candidates}
\end{figure}

As illustrative examples of phosphate-based waste form design, we generate candidates in
the ternary \ce{Na2O-Fe2O3-P2O5} and quaternary \ce{Na2O-Nd2O3-Fe2O3-P2O5} spaces (where
Na/Nd are the waste cations, \ce{Fe2O3} acts as the GFC), using a uniform grid spacing of
2.5\% along each composition axis.
We then apply a combination of domain knowledge-informed rules as well as results from
DFT convex hull analysis to filter out candidate compositions with the following
characteristics (and are therefore unviable):
(1) no salt cations in the composition, 
(2) unreacted elements, unreacted waste salt, or no phosphates in the ground state phase
mixture predicted by the convex hull analysis, and
(3) ground state phase mixtures that contain binary alkali compounds (e.g., \ce{Na3PO4})
that are known to be water soluble.
The resulting filtered list of viable candidate compositions is much smaller (see
\figref{fig:candidates}).

All the relevant target properties for each viable candidate can then be predicted using
the previously trained ML models.
The scoring of candidates, and the identification of candidates to prioritize for
experimental validation and testing (the ``acquisition function''), can be performed one
of several ways depending on the design problem.
For the design of waste forms in particular, it is desirable for the acquisition
function to be:
(1) multiobjective, to simultaneously optimize several target properties such as mass
loading, waste form volume, compressive strength, and so on;
(2) enable end users and domain experts to weight the different targets or success
criteria differently.
For example, Vienna and Kim~\cite{vienna2023preliminary} used a penalty-based approach
to multiattribute optimization of borosilicate glasses:
\begin{equation}\label{eqn:ranking}
  P_i = \left\{
  \begin{aligned}
    w_i \left( \frac{T_i - Y_i}{T_i - L_{i,L}} \right)^n & \quad \text{if $Y_i \leq T_i$} \\
    w_i \left( \frac{Y_i - T_i}{L_{i,U} - T_i} \right)^n & \quad \text{if $Y_i > T_i$}
  \end{aligned}
  \right\}
\end{equation}
where $P_i$ is penalty associated with the $i^{\rm th}$ property, $w_i$ is the $i^{\rm
th}$ property weighting parameter, $T_i$ is the $i^{\rm th}$ property target value,
$Y_i$ is the $i^{\rm th}$ property predicted value, and $L_{i,L}$ and $L_{i,U}$ are the
lower and upper limit for property $i$. 
For each formulation, an optimization is performed by minimizing the sum of penalties
for multiple properties;
(3) able to leverage ML model predictions \textit{as well as prediction uncertainties}
to score candidates.
Acquisition functions traditionally used for Bayesian optimization are well-suited for
this purpose, such as probability of improvement over a specified baseline,
$a_{\rm PI} (x) = \int_{x_0}^{\infty} \mathcal{N}[\mu(x), \sigma^2(x)] dx$, where $x$ is
the target variable to be maximized (e.g., bulk modulus), $x_0$ is the current baseline
performance, and $\mu(x)$ and $\sigma^2(x)$ are the predicted value and uncertainty from
the relevant ML model.
For multiple objectives, a scalarizing function that combines $a_{\rm PI}$ for all the
targets can be defined (e.g., simply a product of $a_{\rm PI}$ values for uncorrelated
targets).

We demonstrate two different strategies for candidate selection in the two example
design spaces.
For the \ce{Na2O-Fe2O3-P2O5} design space, we show a single-objective optimization
targeting salt cation loading in wt\%, which can be estimated from DFT-based convex hull
analysis without the need for ML surrogates.
We choose a candidate composition from among the ones showing the highest loading
($\sim$18\%; higher than the state-of-the-art iron phosphate
glasses~\cite{ebert2019analyses, riley2020dehalogenation, riley2020electrochemical}) for
validation (e.g., ``A2'' in \figref{fig:candidates}a).
For the \ce{Na2O-Nd2O3-Fe2O3-P2O5} design space, we show a multi-objective optimization
with the following three objectives as well as a hard constraint of salt cation loading
$>$20\%:
(1) minimizing melt temperatures ($<$1173~K, the lowest liquidus temperature in the
\ce{Fe2O3-P2O5} binary; see \figref{fig:phase_diagrams}a),
(2) minimizing the waste form volume to immobilize 100~g of salt ($<$100~cm$^3$, lower
than state-of-the-art phosphate glass waste forms~\cite{ebert2019analyses,
riley2020dehalogenation, riley2020electrochemical}), and
(3) maximizing bulk modulus ($>$48~GPa, the highest moduli reported for \ce{Fe2O3-P2O5}
glasses~\cite{chang2003structure, jolley2016iron}).
We use the product of the individual $a_{\rm PI}$ scores (scaled to lie within 0--1) as
the aggregate metric, representing the likelihood of improvement over all three
targets simultaneously (note that the term ``likelihood'' is used loosely, and is not
exactly equivalent to the statistical concept of likelihood), to rank candidates and
choose the most promising ones for experimental synthesis and validation (e.g., ``E1''
and ``E2'' in \figref{fig:candidates}b).

\subsection{Experimental validation}\label{ssec:expt}


The final task in a SL workflow is the validation of selected candidates, and closing
the loop by providing feedback for the refinement of next round of predictions and
candidate selection tasks.
Ideally, the synthesis of the target materials and any post-processing steps should
mimic the final production process as closely as possible, although this is not always
practical (e.g., an industrial process that is hard to replicate in a laboratory).
Similarly, the characterization of the synthesized materials and measurement of
properties should enable direct feedback into the earlier prediction and selection tasks
in the SL workflow;
note that this is not always possible, e.g., properties that are hard to measure
experimentally, mismatch in target performance metrics and properties for which there
exists data to train ML models.

\begin{figure}[!htb]
\centering
\includegraphics[width=\textwidth]{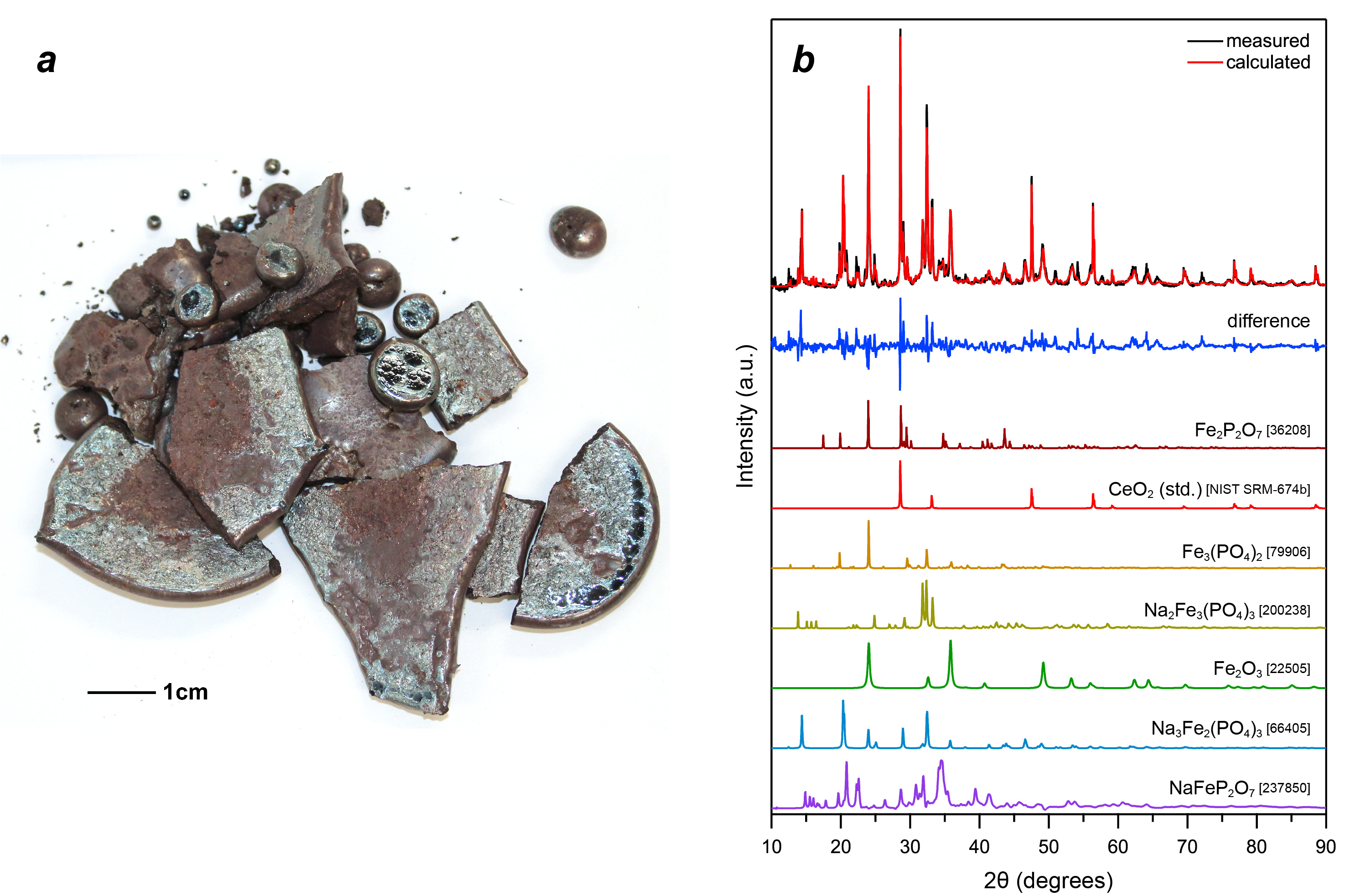}
\caption{
  A2 candidate waste form: (a) picture of the as-quenched sample, (b) Rietveld
  refinement of the diffraction pattern of the as-quenched sample.
  }\label{fig:sample-photo-xrd}
\end{figure}

Here, we synthesize the three selected candidate waste forms (A2, E1, and E2; e.g., see
\figref{fig:sample-photo-xrd}a for an image of a quenched sample) and perform
post-synthesis heat treatment to simulate a canister pour-and-cool for final disposal
(see \secref{ssec:methods-expt} for details of the heat treatment profiles used).
We then characterize the heat-treated samples using X-ray diffraction (e.g., see
\figref{fig:sample-photo-xrd}b for Rietveld refinement for the as-quenched A2
candidate)---determining the phases in each sample and quantifying their distribution
(\tabref{tab:phases}).
We follow up with measurements of candidate properties such as density (i.e., to
validate storage volume estimation) and compressive strength.

{\small
  \begin{xltabular}[!htbp]{\linewidth}{c|c|c|c|c|c|c|c|c}
  \caption{
    Phase distribution (in wt\%) for samples A2, E1, and E2 observed in experiment using
    XRD analysis post heat treatment (SC2 profile for samples A2 and E1, and SC1 profile
    for sample E2; see \secref{ssec:methods-expt}), and the corresponding thermodynamic
    ground state phase distribution predicted by DFT and CALPHAD (sample A2 only).
    R$_{\rm wp}$ is the weighted profile residual for the XRD refinements.
  }\label{tab:phases} \\
  \toprule
  \multirow{2}{*}{\textbf{Phase}} & \multirow{2}{*}{\textbf{ICSD\#}} &
  \multicolumn{3}{c|}{\textbf{Sample A2 [SC2]}} & \multicolumn{2}{c|}{\textbf{Sample E1
  [SC2]}} & \multicolumn{2}{c}{\textbf{Sample E2 [SC1]}} \\
  \cmidrule{3-9}
                     &        &  Expt. & DFT    & CALPHAD & Expt.  & DFT    & Expt.  & DFT    \\
  \midrule
  \ce{NaFe3(PO4)3}   & 61696  &        &        &         &        &        & 33.06  &        \\
  \ce{NaFePO4}       & 193244 &        &        &         &        &        & 30.97  &        \\
  \ce{NaFeP2O7}      & 237850 &        &        &         & 15.96  &        &        &        \\
  \ce{Na3Fe3(PO4)4}  & 95532  &        & 36.70  &         &        &        &        & 36.92  \\
  \ce{Na2Fe3(PO4)3}  & 200238 & 24.41  &        &         & 11.10  &        &        &        \\
  \ce{Na3Fe2(PO4)3}  & 66405  & 23.52  &        &         & 8.10   &        & 6.55   &        \\
  \ce{Na3Fe(PO4)2}   & 85558  &        & 60.92  &         &        & 79.79  &        & 43.98  \\
  \ce{NaNd(PO3)4}    & 401    &        &        &         & 9.80   &        &        &        \\
  \ce{NdPO4}         & 79750  &        &        &         & 10.06  & 20.21  & 11.54  & 19.10  \\
  \ce{NaPO3}         & 174201 &        &        & 11.0    &        &        &        &        \\
  \ce{Na3PO4}        & 33718  & 12.45  &        &         &        &        &        &        \\
  \ce{Na4P2O7}       & 10370  &        &        & 31.0    &        &        &        &        \\
  \ce{Na5P3O10}      & 25837  &        &        & 13.0    &        &        &        &        \\
  \ce{FePO4}         & 79906  & 17.51  &        & 27.0    &        &        &        &        \\
  \ce{Fe2O3}         & 22505  & 9.82   & 2.38   & 18.0    & 8.08   &        & 13.37  &        \\
  \ce{Amorphous}     &        & 12.30  &        &         & 36.90  &        & 4.51   &        \\
  \midrule
  SUM                &        & 100.00 & 100.00 & 100.00  & 100.00 & 100.00 & 100.00 & 100.00 \\
  \midrule
  R$_{\rm wp}$       &        & 6.38   &        &         & 9.29   &        & 7.78   &        \\
  \bottomrule
\end{xltabular}

}

\section{An End-to-End Workflow Run}\label{sec:example-workflow}
To illustrate the end-to-end workflow of an informatics-driven framework, we consolidate
and present one full SL iteration for one of the candidate waste forms discussed
previously, E2, from the \ce{Na2O-Nd2O3-Fe2O3-P2O5} chemical space (with \ce{Na2O} and
\ce{Nd2O3} from the waste stream, \ce{Fe2O3} added as a GFC, and \ce{P2O5} from the
dehalogenation process).

We first generate an enumerated design space of potential candidates in the quaternary
space, as a uniform grid with 2.5~mol\% spacing along each composition axis, for a total
of 12341 candidate compositions.
For each candidate composition, we then apply DFT convex hull analysis to predict the
ground state phase mixture, and apply domain knowledge filters to exclude unviable
compositions, as described in \secref{ssec:candidates}.
The E2 candidate (composition 37.5\% \ce{Na2O} + 5\% \ce{Nd2O3} + 20\% \ce{Fe2O3} +
37.5\% {P2O5} in mol\%) passes all the domain knowledge filters, and a mixture of \{50\%
\ce{Na3Fe(PO4)2} + 21.43\% \ce{Na3Fe3(PO4)4} + 28.57\% \ce{NdPO4}\} (mol\%) is predicted
to be the thermodynamic ground state at its overall composition.
We then estimate 
(a) the salt cation loading (25.29 wt\% of Na, Nd combined),
(b) the waste form storage volume per 100~g of nominal waste salt (an equivalent amount
of \ce{NaCl + NdCl3}) as the the mole fraction weighted DFT-calculated volume of the
ground state phase mixture (54.36~cm$^3$),
(c) target properties (e.g., melt temperature, bulk modulus) using pretrained ML models.
Note that the ML models used here are trained to predict the properties of the
individual phases in the mixture, and the overall properties (and uncertainties) of the
mixture are approximated as the mole fraction-weighted mean.
For the E2 candidate, e.g., the predicted bulk modulus is $75.6 \pm 20.1$, the predicted
melt temperature is $1480.7 \pm 229.3$.
We calculate the probability of improvement for each property (e.g., $a_{\rm PI}(x={\rm
bulk\;modulus}) = \int_{\rm 48 GPa}^{\infty} \mathcal{N}[\mu(x)=75.6, \sigma^2(x)=20.1]
dx = 0.91$) and aggregate the individual scores into a cumulative likelihood of
improvement metric scaled to lie between 0 and 1 (see \figref{fig:candidates}b).

\begin{figure}[!htb]
\centering
\includegraphics[width=0.5\textwidth]{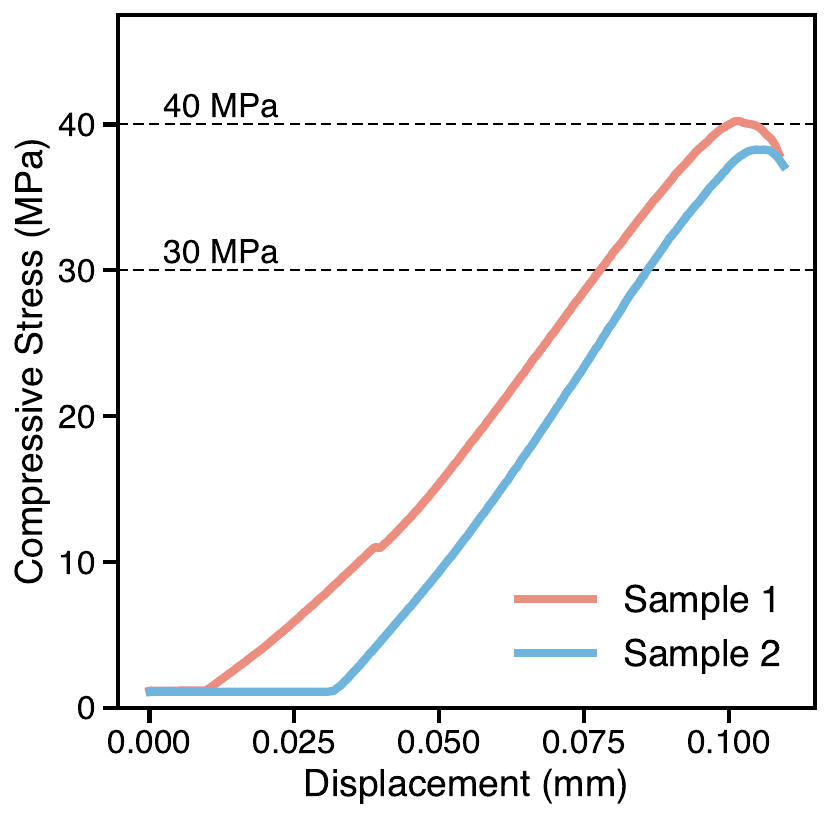}
\caption{
  Compressive strength of the E2 waste form (tested on two different samples).
  }\label{fig:cs}
\end{figure}

We then perform experimental synthesis and heat treatment for the target E2 composition,
and characterize the phases present using X-ray diffraction (see \tabref{tab:phases}).
We measure properties such as density (3.164 g/cm$^3$, and corresponding waste form
volume) and compressive strength (38.3--40.2~GPa; see \figref{fig:cs}), with all
properties surpassing the target performance criteria listed at the end of
\secref{sec:prior-art}.
The characterization data and measured properties are used to refine the outputs from
the physics-based modeling tasks, as well as to augment existing training data and
retrain ML models for the next SL iteration; these efforts will be reported in the
future.

\section{Discussion and Outlook}\label{sec:discussion}
While such as informatics-driven workflows are powerful tools for accelerating materials
design in general, and waste form design in particular, some challenges remain and we
discuss a few salient ones below.

The prediction of the exact phases formed during synthesis (and their respective
distributions) given the overall composition and a set of experimental conditions and
post-processing steps remains an open question.
Analyses based on DFT convex hull and related thermodynamic stability measures, while
still extremely useful, suffer from limitations related to DFT accuracy and ideal
conditions (zero temperature and, often, ambient pressure).
As seen from \tabref{tab:phases}, the mismatch between computational predictions and
experimental realization is large.
Some phases that have been reported experimentally are predicted to be 0~K
unstable by DFT (e.g., \ce{Na2Fe3(PO4)3} found in both A2 and E1 candidates is predicted
to be 60~meV/atom above the convex hull of formation energy).
Some other phases are excluded from the DFT analyses due the nature of the problem set
up; i.e., with the overall melt modeled as a mixture of the corresponding oxides,
certain compounds with mixed elemental oxidation states (e.g., \ce{NaFe3(PO4)3} found
in the E2 candidate, with two nominal \ce{Fe^{3+}} and one nominal \ce{Fe^{2+}}) or
oxidation states different from the one in the melt ingredient (e.g., \ce{NaFePO4} with
Fe$^{2+}$, in contrast to the GFC additive \ce{Fe2O3} in the melt) are naturally
excluded.
The latter problem can be mitigated by performing DFT convex hull analyses with varying
elemental chemical potential (e.g., $\mu_{\rm O}$) but increases the computational
expense and complexity of the task.
While CALPHAD approaches can consider effects of variables such as temperature and
chemical potential is calculating phase diagrams, they are severely limited by the lack
of assessed databases, especially for compositions relevant to phosphate-based waste
forms (e.g., current CALPHAD databases in FactSage do not include thermodynamic data for
any quaternary phases in the \ce{Na-Fe-P-O} chemical space [resulting in the mismatch
for sample A2 in \tabref{tab:phases}], and assessed databases for the \ce{Na-Nd-Fe-P-O}
chemical space do not exist).
Further, while recent work has shown the possibility of predicting crystallization
pathways from amorphous matter using a combination of ab-initio methods and deep
learning~\cite{aykol2023predicting}, further efforts are needed to develop a holistic
framework that fully bridges the gap between computational design and experimental
realization of materials.

Another general challenge is the lack of large, well-curated datasets to train ML models
to predict several waste form-relevant properties of interest.
For example, properties such as compressive strength are not widely reported for a large
set of materials and are not trivial to calculate using ab-initio techniques.
Similarly, chemical durability measurements of ceramic waste forms are scattered across
the literature and no centralized curated dataset exists (in contrast to traditional
borosilicate glass-based waste forms, e.g.).
While some of these challenges may be mitigated by using domain knowledge integration
(DKI; see below) or using techniques such as transfer
learning~\cite{hutchinson2017overcoming, yamada2019predicting, jha2019enhancing,
cubuk2019screening, gupta2021cross, chen2021atomsets}, we urge for investments into
large-scale data collection and curation of waste form properties that can benefit the
community as a whole.

For effective use of informatics-based approaches for the design of waste forms,
especially in the context of the above-mentioned challenges, it is crucial to leverage
DKI from experts in various parts of the SL workflow.
Such DKI can take many forms:
(1) identifying
data-abundant waste form properties that are correlated with data-scarce performance
metrics (e.g., cohesive energy to inform chemical durability, viscosity and liquidus
temperature to inform processability, bulk modulus to compressive strength [as used in
this work]),
(2) defining and constraining the design spaces of interest (e.g., identifying the
correct target alkali and rare-earth waste cations, and any potential pool of additives
that can be used in waste form fabrication),
(3) filtering undesirable candidates (e.g., excluding candidate waste form compositions
that are expected to form water-soluble alkali phosphates), and so on.

Overall, the proposed approach of using informatics-based approaches for waste form
design has the potential to drastically reshape the way that these types of efforts are
conducted across the world.
The largest impact of this method, in contrast to the Edisonian trial and error
approach, is the potential to cut the time required to find optimal solutions by an
order of magnitude (or more).
Tapping into crystallographic databases, thermodynamic property databases, phase
diagrams, as well as other material property databases, this informatics approach can be
used to design waste forms containing anywhere from high amorphous (glassy) phase
fractions to high crystalline fractions as well as mixtures thereof.
Finally, we note that the opportunities afforded by such approaches extend far past
borosilicate and phosphate waste forms, and these techniques can be used for optimize
the processing history and conditions to effectively fabricate the next generation of
waste forms as well as aiding eventual scale-up efforts for promising waste forms once
so identified.
Some challenges related to bridging the gap between predictions and laboratory
realization of waste forms exist, but current approaches already present avenues for
significant acceleration of the design of novel waste forms.


\section{Methods}\label{sec:methods}
\subsection{Thermodynamic analysis}\label{ssec:methods-thermo}
All DFT-based convex hull analysis was performed using data openly available from the
Materials Project~\cite{jain2013commentary} and the pymatgen~\cite{ong2013python}
software package.

All CALPHAD calculations were performed using the commercial FactSage
software~\cite{bale2002factsage} and associated databases.

\subsection{Machine learning}\label{ssec:methods-ml}
All machine learning models reported in this work were built using the open-source
lolo~\cite{lolo} random forests library, using default hyperparameters, trained on open
datasets listed in \tabref{tab:datasets} (i.e., those for melting temperature and bulk
modulus).
The additional \ce{Fe2O3-P2O5} liquidus data used in \figref{fig:ml} was calculated
using FactSage, and the data is made available.

\subsection{Experimental synthesis and characterization}\label{ssec:methods-expt}

The raw reagents (i.e., \ce{Na2CO3} [Aldrich, 99.9\% trace metals], \ce{Nd2O3},
\ce{Fe2O3} [Baker, 100.4\%], and \ce{NH4H2PO4} [Sigma Aldrich, $\geq$98.5\%]) were
batched using an analytical balance (Mettler Toledo ME204).
The reagents were then loaded into 250~mL alumina crucibles (ACC3742, McDanel Advanced
Ceramic Technologies) with a Pt/10\%Rh lid and melted in a high-temperature furnace
(Deltech Furnaces, Inc.) for 2~hour delay at 1250\degc, with a ramp heating rate of
5\degc/min, followed by quenching on an Inconel plate.
The quenched materials were ground to a fine particle size in a tungsten carbide milling
chamber and run through slow cooling process with an aliquot of the ground material
within a 10~mL alumina crucible (ACM3760, McDanel Advanced Ceramic Technologies).

Two custom heat treatment profiles were employed: ``SC2'' for samples A2 and E1, and
``SC1'' for sample E2.
SC2 included a ramp rate of 5\degc/min from room temperature to 1250\degc, a dwell for 1
hour at 1250\degc, a $-5$\degc/min cooling to 500\degc, a reheat at 5\degc/min to
600\degc, a dwell for 12 hours, a ramp cool at $-0.03$\degc/min to 465\degc over the
course of 75 hours, and ended with a $-1$\degc/min ramp cooling rate.
SC1 included a ramp rate of 5\degc/min from room temperature to 1250\degc at
5\degc/minute, held for 1 hour, ramp cooled at $-25$\degc/minute to 1000\degc, and then
ramp cooled to room temperature at $-0.1^{\circ}$C/minute.
These custom heat treatments were designed to homogenize the melt at 1250$^{\circ}$C,
rapidly cool to above the glass transition temperature ($T_g$), reheat and then cool
slowly to initiate crystallization.

Once the samples were cooled, vertical slides were prepared using glycol suspensions and
a slow speed Buehler diamond saw. 
The samples were removed from their alumina crucibles, ground in a tungsten carbide
milling chamber to a fine particle size and analyzed using X-ray diffraction.
Then, a known amount (5 mass\%) \ce{CeO2} of a NIST Standard Reference Material
(SRM-674b) was added and ground for 30 additional seconds.
Samples doped with \ce{CeO2} were run with a Bruker D8 Advance diffractometer in a scan
range of 5--90$^{\circ}$ 2$\theta$, with a 0.01486$^{\circ}$ 2$\theta$ step angle, with
1 second dwells per step.
The diffraction patterns were analyzed using Bruker Topas (version 5) software with
PDF5+ International Centre for Diffraction Data (ICDD) and Inorganic Crystal Structure
Database (ICSD).

Compressive strength tests were run in duplicate on an Instron 5582 (ID 5582R1924) using
a fixed rate of 0.1~mm/minute in accordance with ASTM C1358-18.
Samples were prepared in a 2:1 geometry of height:width using a series of procedures
from a diamond wire saw, a slow-speed diamond blade saw, and polishing processes.
The preferred sample geometry is a cylinder but making cylinders from these samples
proved difficult due to small sample size.
Thus, rectangular prisms were made that were on the order of $\sim$6~mm tall by
$\sim$3~mm wide.
Samples were loaded until failure and the compressive strength was reported as the
maximum uniaxial compressive stress reached when the material failed.

\clearpage

\section*{Data Availability}
All associated research data has been deposited into a repository on
figshare~\cite{figshare2024towards}.

\section*{Author Contributions}
Conceptualization: VIH, JES, BJR, JDV, JD;
Data curation: VIH, MP, SIA, XL, JO, RJS, EKN, BJR, JDV, TM, JK, TN;
Formal anaylsis: VIH, MP, JO, RJS, EKN, BJR, JDV, TM, JK, TN;
Funding acquisition: VIH, JES, BJR, JDV, JD;
Investigation: VIH, MP, JO, RJS, EKN, BJR, JD, TM, JK, TN;
Methodology: VIH, SIA, JES, MP, XL, RJS, EKN, BJR, JDV, TM, JK, TN;
Project administration: JES, JDV, JD;
Resources: BJR, JDV, JD;
Software: VIH, SIA;
Supervision: JES, BJR, JDV, JD;
Validation: VIH, BJR, TM;
Visualization: VIH, MP, BJR, JD, TM, JK, TN;
Writing -- original draft: VIH, SIA, JES, MP, XL, BJR, JDV, JD, TM;
Writing -- review and editing: all authors.

\section*{Conflicts of Interest}
There are no conflicts of interest to declare.

\section*{Acknowledgments}
The information, data, or work presented herein was funded in part by the Advanced
Research Projects Agency-Energy (ARPA-E), U.S.\ Department of Energy, under Award Number
DE-AR0001613.
The views and opinions of authors expressed herein do not necessarily state or reflect
those of the United States Government or any agency thereof.
Pacific Northwest National Laboratory (PNNL) is operated by Battelle Memorial Institute
for the DOE under contract DE-AC05-76RL01830.
The authors thank Anthony Guzman (PNNL), Irving L.\ Brown (PNNL), and Timothy J.\
Roosendaal (PNNL) for help with preparing samples for compressive strength, which proved
to be quite challenging, and Jarrod V. Crum (PNNL) for reviewing the manuscript.

\begin{spacing}{0.9}
\printbibliography
\end{spacing}

\end{document}